\documentclass[12pt]{iopart}
\usepackage{amssymb}
\usepackage{bm}

\renewcommand{\vec}[1]{\bm{#1}}
\newcommand{\uvec}[1]{\hat{\vec{#1}}}

\newcommand{\avr}[1]{\left\langle#1\right\rangle}
\newcommand{\bra}[1]{\left\langle#1\right|}
\newcommand{\ket}[1]{\left|#1\right\rangle}

\newcommand{\Lv}{\mathcal{L}}
\newcommand{\U}{\mathsf{U}}
\renewcommand{\P}{\mathcal{P}}
\newcommand{\Q}{\mathcal{Q}}

\newcommand{\LT}{\textsf{LT}}

\begin{document}

\title{A classical long-time tail in a driven granular fluid}
\author{W T Kranz}
\address{Georg-August-Universit\"at G\"ottingen, Institut f\"ur
  Theoretische Physik, Friedrich-Hund-Platz 1, 37077 G\"ottingen,
  Germany}
\address{Max-Planck-Institut f\"ur Dynamik und Selbstorganisation, Am
  Fa\ss berg 17, 37077 G\"ottingen, Germany}
\ead{kranz@theorie.physik.uni-goettingen.de}

\begin{abstract}
  I derive a mode-coupling theory for the velocity autocorrelation
  function, $\psi(t)$, in a fluid of randomly driven inelastic hard
  spheres far from equilibrium. With this, I confirm a conjecture from
  simulations that the velocity autocorrelation function decays
  algebraically, $\psi(t)\propto t^{-3/2}$, if momentum is
  conserved. I show that the slow decay is due to the coupling to
  transverse currents.
\end{abstract}
\submitto{\JPCM}

\section{Introduction}
\label{sec:introduction}

The algebraic, rather than exponential in time decay of the velocity
autocorrelation function (VACF), $\psi(t)\propto t^{-\alpha}$, in
simple fluids had been quite a surprise when it was first discovered
\cite{alder+wainwright67,alder+wainwright70a}. It was finally
explained by mode coupling theories and attributed to vortex flows
\cite{pomeau+resibois75}. Long-time tails are expected even in high
energy physics \cite{kovtun+yaffe03} now and have been reported
recently also for fluids far from equilibrium \cite{golestanian09}. In
particular they are discussed for granular fluids
\cite{ahmad+puri07,hayakawa+otsuki07,otsuki+hayakawa09a,orpe+kumaran08,kumaran09,fiege+aspelmeier09}.

Here, I show that the original mode coupling argument
\cite{resibois75,furtado+mazenko76,bosse+goetze78} can be adapted to
the stationary state of a randomly driven granular fluid. In
particular, I explain the observation that $\alpha\approx 1.5$ in
three space dimensions \cite{fiege+aspelmeier09}. This is exactly the
same exponent as for equilibrium fluids and stands in contrast to a
number of unconventional exponents reported in the literature.

In a fluid in thermal equilibrium, long-time tails are a result of the
coupling to the transverse current modes, $\vec j_{\vec
  k}^{\mathrm{T}}$, labelled by the wave vector $\vec k$. A number of
approaches (see \cite{pomeau+resibois75} and references therein, and
\cite{furtado+mazenko76,bosse+goetze78,cukier+mehaffey78,sjoegren+sjoelander79,kirkpatrick+nieuwoudt86})
confirmed the result $\psi(t\to\infty) \propto t^{-3/2}$. In a Lorentz
gas, momentum is not conserved and it was argued
\cite{ernst+weyland71} that this leads to a faster decay,
$\psi(t\to\infty) \propto t^{-5/2}$. See \cite{hoefling+franosch07}
for why this behaviour may be hard to observe.

For a freely cooling granular gas, a long-time tail in the number of
collisions, $\tau$, is predicted of the form $\psi(\tau\to\infty)
\propto \tau^{-3/2}$ \cite{hayakawa+otsuki07}. Here, the coupling to
the longitudinal and transverse current are both relevant. For
shear-driven granular fluids, there are two competing
proposals. Hayakawa and Otsuki \cite{otsuki+hayakawa09a} predict
$\psi(t\to\infty) \propto (\dot\gamma t)^{-5/2}$, where $\dot\gamma$
is the shear rate and Kumaran \cite{kumaran09} predicts
$\psi(t\to\infty) \propto (\dot\gamma t)^{-7/2}$ in the vorticity
direction and a slightly faster decay, $\psi(t\to\infty) \propto
(\dot\gamma t)^{-15/4}$, in the gradient and flow directions. The
difference remains unresolved \cite{otsuki+hayakawa09b}. In both
theories, the physical interpretation of the relevant collective modes
is not obvious.

From the above discussion one can conclude that the existence of
long-time tails seems to be rather universal even in fluids far from
equilibrium. Two questions, however, have to be answered for every
specific system: What is the mechanism that induces the slow decay and
what is the value of the exponent $\alpha$? In the following, I will
address these two questions for the randomly driven granular fluid.

The paper is organised as follows. We start in section~\ref{sec:model}
by defining our model system. In section~\ref{sec:micr-descr} I give
the formally exact equation of motion for the VACF. This will be
closed in section~\ref{sec:mode-coupl-appr} by a mode-coupling
approximation. In section~\ref{sec:discussion} I discuss the results
of the approximation. In particular, the long-time tails. In the final
section~\ref{sec:concl--persp} I summarise my results and give some
perspectives for future work.

\section{Model}
\label{sec:model}

\subsection{Inelastic Hard Spheres}
\label{sec:inel-hard-spher}

The granular fluid is modelled as a monodisperse system of $N$ smooth
inelastic hard spheres of diameter $d$ and mass $m=1$ in a volume
$V=L^3$. I consider the thermodynamic limit $N,V\to\infty$ such that
the density $n=N/V$ remains finite. Dissipation is introduced through
a constant coefficient of normal restitution $\varepsilon\in[0,1]$
that augments the law of reflection \cite{haff83},
\begin{equation}
  \label{eq:1}
  \uvec r_{12}\cdot\vec v'_{12} = -\varepsilon\,\uvec r_{12}\cdot\vec v_{12},
\end{equation}
where $\vec v_{12} = \vec v_1 - \vec v_2$ is the relative velocity and
$\uvec r_{12}$ is the unit vector pointing from the centre of particle
2 to particle 1. The prime indicates post-collisional quantities.

\subsection{Stochastic Driving Force}
\label{sec:stoch-driv-force}

The driving force is implemented as an external random force,
\begin{equation}
  \label{eq:2}
  \vec v'_i(t) = \vec v_i(t) + \sqrt{P_{\mathrm D}}\,\vec\xi_i(t),
\end{equation}
where $P_{\mathrm D}$ is the driving power. The $\xi_i^{\alpha}$, $\alpha=x,y,z$
are Gaussian random variables with zero mean and variance,
\begin{equation}
  \label{eq:3}
  \avr{\xi_i^{\alpha}(t)\xi_j^{\beta}(t')}_{\xi} 
  = [\delta_{ij} - \delta_{\pi(i),j}]\delta^{\alpha\beta}\delta(t-t'),
\end{equation}
where $\pi(i)$ denotes the nearest neighbour of particle $i$. In
effect, the two particles $i$ and $\pi(i)$ are driven by forces of
equal strength but opposite direction. Thereby, the external force
does not destroy momentum conservation on macroscopic length scales
\cite{fiege+aspelmeier09,espanol+warren95}. 

\subsection{The Granular Fluid}
\label{sec:granular-fluid}

Macroscopically, the fluid is fully characterised by the packing
fraction, $\varphi = \pi nd^3/6$, the coefficient of restitution,
$\varepsilon$, and the driving power, $P_{\mathrm D}$.  In the stationary state,
the granular temperature $T = T(\varphi,\varepsilon,P_{\mathrm D}) =
\frac{1}{3N}\sum_i\vec v_i^2$ is given by the balance between the
driving power, $P_{\mathrm D}$, and the energy loss through the inelastic
collisions.

The collision frequency $\omega_c\propto\sqrt T$ is the only time
scale of the system. Thus, changing the granular temperature only
changes the time scale of the system. I use this freedom and set
$T\equiv1$ in the following.

\section{Microscopic Description}
\label{sec:micr-descr}

\subsection{Phase Space Distribution}
\label{sec:phase-space-distr}

In contrast to fluids in thermal equilibrium, no analytical expression
for the stationary phase space distribution of driven granular fluids
is known so far. Therefore, I have to make a few assumptions to
evaluate the expectation values. First of all I assume that positions
and velocities are uncorrelated, $\varrho(\Gamma) = \varrho_{\mathrm
  r}(\{\vec r_i\})\varrho_{\mathrm v}(\{\vec v_i\})$. Moreover, I
assume that the velocity distribution factorises into a product of one
particle distribution functions, $\varrho_{\mathrm v}(\{\vec v_i\}) =
\prod_i\varrho_1(\vec v_i)$. All we need to know about $\varrho_1(\vec
v)$ are a few moments. Namely, that it has a vanishing first moment,
$\int\rmd^3v\,\vec v\varrho_1(\vec v) = 0$, a finite second moment,
$\int\rmd^3v\,v^2\varrho_1(\vec v) = 3T < \infty$ and a finite third
collisional moment, $\int\rmd^3v(\uvec r\cdot\vec v)^3\Theta(-\uvec
r\cdot\vec v)\varrho_1(\vec v) < \infty$. The spatial
distribution function, $\varrho_{\mathrm r}(\{\vec r_i\})$, enters the
theory via static correlation function, as will be discussed below.

Averages over pairs of observables define a scalar product, $\avr{A|B}
:= \avr{A^*B} := \int\rmd\Gamma\varrho(\Gamma)A^*(\Gamma)B(\Gamma)$,
where $A^*$ denotes the complex conjugate of $A$.

\subsection{Observables}
\label{sec:observables}

The VACF, $\psi(t) = \avr{\vec v_s|\vec v_s(t)}/3$, is defined in
terms of the tagged particle velocity $\vec v_s$. The tagged particle
position will be described by the density $\rho^s(\vec r,t) =
\delta(\vec r_s-\vec r(t))$. The host fluid is characterised by the
density and current fields \numparts
\begin{eqnarray}
  \label{eq:7}
  \rho(\vec r, t) &= \frac1N\sum_i\delta(\vec r - \vec r_i(t)),\\
  \vec j(\vec r, t) &= \frac1N\sum_i\vec v_i(t)\delta(\vec r - \vec r_i(t)).
\end{eqnarray}
\endnumparts
In particular, I will use the spatial Fourier transform of those
fields, $\rho_{\vec k}^s(t), \rho_{\vec k}(t)$, and the longitudinal
and transverse current fields $j_{\vec k}^L(t) = \uvec k\cdot\vec
j_{\vec k}(t)$, and $\vec j_{\vec k}^T(t) = \vec j_{\vec k}(t) - \uvec
kj_{\vec k}^L(t)$, respectively.

\subsection{Dynamics}
\label{sec:dynamics}

We have shown in \cite{kranz+sperl13} that the time evolution operator
$\U(t) = \exp(\rmi t\Lv_+)$ can be written in terms of an effective
pseudo Liouville operator $\Lv_+$ \cite{altenberger75}. It is given as
a sum of three parts, $ \Lv_+ = \Lv_0 + \mathcal T_+ + \Lv_{\mathrm
  D}^+$, which are in turn: The free streaming operator $\Lv_0$, the
collision operator $\mathcal T_+$, and the driving operator
$\Lv_{\mathrm D}^+$.

With the Mori projectors $\P = \ket{\vec v_s}\bra{\vec v_s}/3$, $\Q =
1 - \P$, one can derive a formally exact equation of motion for the
VACF
\begin{equation}
  \label{eq:6}
  \dot\psi(t) + \frac{1+\varepsilon}{3}\omega_{\mathrm E}\psi(t) 
  + \omega_{\mathrm{E}}^2\int_0^t\rmd\tau m(t-\tau)\psi(\tau) = 0,
\end{equation}
where the local term $\avr{\vec v_s|i\Lv_+\vec v_s}/3 =
-(1+\varepsilon)\omega_{\mathrm{E}}/3$ was determined in
\cite{fiege+aspelmeier09}. The memory kernel is formally given as
\begin{equation}
  \label{eq:7}
  m(t) = \avr{\vec v_s|\Lv_+\Q\tilde\U(t)\Q\Lv_+\vec v_s}/3\omega_{\mathrm{E}}^2
\end{equation}
and $\tilde\U(t) = \exp(\rmi t\Q\Lv_+\Q)$ is a modified propagator
\cite{boon+yip92,mori65b,kranz+sperl13}. The Enskog collision
frequency $\omega_{\mathrm E} = 24\varphi\chi/\sqrt\pi d$ is given in
terms of the contact value of the pair correlation function at
contact, $\chi$ \cite{hansen+mcdonald06}.

\section{Mode-Coupling Approximations}
\label{sec:mode-coupl-appr}

I consider three contributions to the memory kernel $m(t) \approx
m_{\rho}(t) + m_{\mathrm L}(t) + m_{\mathrm{T}}(t)$ that are induced
by the coupling of the tagged particle to the host fluid. Namely, to
the collective density field [$m_{\rho}(t)$], and to the longitudinal
and transverse current field [$m_{\mathrm L}(t)$ and
$m_{\mathrm{T}}(t)$, respectively].

The behaviour of the collective modes is characterised by their
two-point correlation functions, 
\numparts
\begin{eqnarray}
  \label{eq:15}
  \phi(k, t) &= N\avr{\rho_{\vec k}|\rho_{\vec k}(t)}/S_k,\\
  \phi_{\mathrm L}(k, t) &= N\avr{j_{\vec k}^{\mathrm{L}}|j_{\vec k}^{\mathrm L}(t)},\\
  \phi_{\mathrm{T}}^{\alpha\beta}(k, t) 
  &= N\avr{j_{\vec k}^{\mathrm{T}\alpha}|j_{\vec k}^{\mathrm{T}\beta}(t)}
  = \phi_{\mathrm{T}}(k, t)\delta^{\alpha\beta}, 
\end{eqnarray}
\endnumparts
where $S_k = N\avr{\rho_{\vec k}|\rho_{\vec k}}$ is the static
structure factor, and
\begin{equation}
  \label{eq:18}
  \phi_{\mathrm{s}}(k, t) = \avr{\rho_{\vec k}^{\mathrm{s}}|\rho_{\vec k}^{\mathrm{s}}(t)}
\end{equation}
is the incoherent scattering function.

In terms of these correlation functions, I replace the modified
propagator
\begin{eqnarray}
  \label{eq:17}
  \eqalign{
  \tilde\U(t) &\approx N\sum_{\vec k}\ket{\rho_{\vec k}\rho_{-\vec k}^{\mathrm s}}
  \phi(k, t)\phi_{\mathrm s}(k, t)\bra{\rho_{\vec k}\rho_{-\vec k}^{\mathrm s}}/S_k\\
  &+ N\sum_{\vec k}\ket{j_{\vec k}^L\rho_{-\vec k}^{\mathrm s}}
  \phi_{\mathrm L}(k, t)\phi_{\mathrm s}(k, t)\bra{j_{\vec k}^L\rho_{-\vec k}^{\mathrm s}}\\
  &+ \frac N2\sum_{\vec k}\ket{\vec j_{\vec k}^{\mathrm{T}}\rho_{-\vec k}^{\mathrm s}}
  \phi_{\mathrm T}(k, t)\phi_{\mathrm s}(k, t)\bra{j_{\vec k}^{\mathrm{T}}\rho_{-\vec k}^{\mathrm s}}}
\end{eqnarray}
by a mode-coupling approximation. Similar approximations have been
made in, e.g., \cite{resibois75,furtado+mazenko76}.

The coupling to the collective density field then reads
\begin{equation}
  \label{eq:19}
  3\omega_{\mathrm{E}}^2m_{\rho}(t) 
  = \sum_{\vec k}\mathcal V_{\vec k}^{\rho}\mathcal W_{\vec k}^{\rho}
  \phi(k,t)\phi_{\mathrm s}(k,t),
\end{equation}
where the vertices
\numparts
\begin{eqnarray}
  \label{eq:20}
  \mathcal V_{\vec k}^{\rho} &= \sqrt{N/S_k}
  \avr{\vec v_s|\Lv_+\Q\rho_{\vec k}\rho_{-\vec k}^{\mathrm s}} = \vec k(S_k-1)/\sqrt{NS_k},\\
  \mathcal W_{\vec k}^{\rho} &= \sqrt{N/S_k}
  \avr{\rho_{\vec k}\rho_{-\vec k}^{\mathrm s}|\Q\Lv_+\vec v_s} =
  \frac{1+\varepsilon}{2}\vec k(S_k-1)/\sqrt{NS_k} 
\end{eqnarray}
\endnumparts
can be deduced from equations (46) and (47) in
\cite{kranz+sperl13}. Explicitly, we find
\begin{equation}
  \label{eq:21}
  \omega_{\mathrm{E}}^2m_{\rho}(t) 
  = \frac{2\pi^2}{9}\frac{1+\varepsilon}{2}\frac{d^3}{\varphi}
  \int_0^{\infty}\!\frac{\rmd kk^4}{(2\pi)^3}S_k(nc_k)^2
  \phi(k,t)\phi_{\mathrm s}(k,t),
\end{equation}
where $nc_k = 1 - 1/S_k$ is the direct correlation function
\cite{hansen+mcdonald06}. This implies that
$\omega_{\mathrm{E}}^2m_{\rho}(t) \equiv m_0(t)$, where $m_0(t)$ is
given in \cite{sperl+kranz12} as the memory kernel for the mean square
displacement. In \cite{sperl+kranz12} we were concerned with the
behaviour at high densities close to the glass transition and we used
a mode-coupling approximation for the coherent scattering function,
$\phi(k,t)$, itself. Here, I am interested in the regime of moderate
densities, instead. Consequently, below I will use a hydrodynamic
expression for the coherent scattering function
[equation~(\ref{eq:28})].

The coupling to the currents reads
\numparts
\begin{eqnarray}
  \label{eq:22}
  3\omega_{\mathrm{E}}^2m_{\mathrm L}(t)  &= \sum_{\vec k}
  \mathcal V_{\vec k}^L\mathcal W_{\vec k}^L
  \phi_{\mathrm L}(k, t)\phi_{\mathrm s}(k, t),\\
  3\omega_{\mathrm{E}}^2m_{\mathrm T}(t) &= \frac12\sum_{\alpha,\beta}\sum_{\vec k}
  \mathcal V_{\vec k}^{\alpha\beta}\mathcal W_{\vec k}^{\alpha\beta}
  \phi_{\mathrm T}(k,t)\phi_{\mathrm s}(k,t),
\end{eqnarray}
\endnumparts
where the vertices
\numparts
\begin{eqnarray}
  \label{eq:23}
  \mathcal V_{\vec k}^{\mathrm{L}} &= \sqrt N
  \avr{\vec v_s|\Lv_+\Q j_{\vec k}^L\rho_{-\vec k}^{\mathrm s}},\\
  \mathcal W_{\vec k}^{\mathrm{L}} &= \sqrt N
  \avr{j_{\vec k}^L\rho_{-\vec k}^{\mathrm s}|\Q\Lv_+\vec v_s},
\end{eqnarray}
\endnumparts
and
\numparts
\begin{eqnarray}
  \label{eq:24}
  \mathcal V_{\vec k}^{\alpha\beta} &= \sqrt N
  \avr{v_s^{\beta}|\Lv_+\Q j_{-\vec k}^{T\alpha}\rho_{\vec k}^{\mathrm s}},\\
  \mathcal W_{\vec k}^{\alpha\beta} &= \sqrt N
  \avr{j_{-\vec k}^{T\alpha}\rho_{\vec k}^{\mathrm s}|\Q\Lv_+v_s^{\beta}}
\end{eqnarray}
\endnumparts
are calculated in \ref{sec:appendix}. While $\mathcal V_{\vec
  k}^{\rho} \ne \mathcal W_{\vec k}^{\rho}$ indicates the violation of
time reversal invariance in the dissipative fluid, one finds 
\numparts
\begin{eqnarray}
  \label{eq:25}
  \mathcal V_{\vec k}^{\mathrm{L}} = \mathcal W_{\vec k}^{\mathrm{L}} 
  &= \rmi\frac{1+\varepsilon}{3}\uvec k\omega_{\mathrm{E}}U_{\mathrm{L}}(kd)/\sqrt N,\\
  \label{eq:12}
  \mathcal V_{\vec k}^{\alpha\beta} = \mathcal W_{\vec k}^{\alpha\beta}
  &= \rmi\sqrt{2/3}\delta^{\alpha\beta}
  \frac{1+\varepsilon}{3}\omega_{\mathrm{E}}U_{\mathrm{T}}(kd)/\sqrt N,
\end{eqnarray}
\endnumparts
where $U_{\mathrm{L}}(x) = 3j''_0(x)$ and $U_{\mathrm{T}}(x) =
\sqrt6j'_0(x)/x$ are effective potentials. Here, $j_0(x)$ is the
zeroth order spherical Bessel function \cite{gradshteyn00} and the
prime denotes the derivative with respect to the argument. Notably,
the effective potentials are independent of density. The vertices are
similar in form to those found in \cite{resibois75,furtado+mazenko76}.

For the memory kernels, we find\footnote{The apparent divergence for
  $\varphi\to0$ is spurious as $\omega_{\mathrm{E}} \sim
  \Or(\varphi)$.}
\begin{equation}
  \label{eq:26}
  m_{\mathrm{L},\mathrm{T}}(t) 
  = -\frac{8\pi^2}{81}\frac{(1+\varepsilon)^2}{4}\frac{d^3}{\varphi}
  \int_0^{\infty}\!\frac{\rmd kk^2}{(2\pi)^3}U^2_{\mathrm{L},\mathrm{T}}(kd)
  \phi_{\mathrm{L},\mathrm{T}}(k, t)\phi_{\mathrm s}(k, t).
\end{equation}
Given a static structure factor, $S_k$, and the dynamic
correlator~(\ref{eq:15},b,c) and (\ref{eq:18}) the approximate memory
kernel is fully determined by equations (\ref{eq:21}) and
(\ref{eq:26}). All three contributions to the approximate memory
kernel diverge in the short time limit. Actually, the memory kernel
should vanish for $t\to0$. For elastic hard spheres, a number of
proposals to that end have been made
\cite{resibois75,furtado+mazenko76,cukier+mehaffey78}. As I am only
interested in the asymptotic behaviour, I will not further discuss
this divergence.

\section{Discussion}
\label{sec:discussion}

The long-time asymptotics, $\psi(t\to\infty)$, are related to the
limit $\lim_{s\to0}s\hat\psi(s)$ in the Laplace domain\footnote{I use
  the convention $\hat f(s) = \LT[f](s) =
  \rmi\int_0^{\infty}f(t)\rme^{-\rmi st}\rmd t$.}. For small $s$ we
have
\begin{eqnarray}
  \label{eq:30}
  s\hat\psi(s) 
  &= s[-\rmi\omega_{\mathrm{E}} + s - \omega_{\mathrm{E}}^2\hat m(s)]^{-1}\nonumber\\
  &\simeq \rmi\frac{s}{\omega_{\mathrm{E}}} + \frac{s^2}{\omega_E^2} - s\hat m(s),
\end{eqnarray}
i.e., $\lim_{s\to0}s\hat\psi(s) = -\lim_{s\to0}s\hat m(s)$ or
\begin{equation}
  \label{eq:37}
  \psi(t\to\infty) = -m(t\to\infty).
\end{equation}
The long-time tails of the VACF are identical (up to the sign) to
those of the associated memory kernel.

At moderate densities, a driven granular fluid is well described by
Navier-Stokes order hydrodynamic equations
\cite{goldhirsch03,vollmayr+aspelmeier11}. Consequently, I assume that
the dynamic correlation functions take the following form: 
\numparts
\begin{equation}
  \label{eq:27}
  \phi_{\mathrm s}(k,t) = \rme^{-Dk^2t},
\end{equation}
where $D$ is the diffusion coefficient,
\begin{equation}
  \label{eq:28}
  \phi(k,t) = \cos(ckt)\rme^{-\Gamma k^2t},\quad
  \phi_{\mathrm L}(k,t) = \ddot\phi(k,t)/k^2,
\end{equation}
where $c$ is the speed of sound and $\Gamma$ the sound damping
constant, and,
\begin{equation}
  \label{eq:29}
  \phi_{\mathrm T}(k,t) = \rme^{-\eta k^2t}
\end{equation}
\endnumparts
with the shear viscosity $\eta$. 

All the transport coefficients and the speed of sound are functions of
the coefficient of restitution $\varepsilon$. For the diffusion
coefficient, Fiege \textit{et al.}  \cite{fiege+aspelmeier09} found
$D(\varepsilon) \propto 2/(1 + \varepsilon)$. According to van Noije
\textit{et al.}  \cite{vannoije+ernst99} the sound damping constant is
given as $\Gamma = \nu + D_{\Gamma}$ where $\nu$ is the kinematic
viscosity and $D_{\Gamma}(\varepsilon)\propto 1/(1 - \varepsilon^2)$ is a
term peculiar to inelastic fluids. The viscosities $\eta$ and $\nu$
have a more complicated dependence on the degree of dissipation
\cite{garzo+montanero02}. The speed of sound, $c$, is smaller in a
fluid of inelastic compared to elastic hard spheres but only weakly
depends on the value of the coefficient of restitution, $\varepsilon$
\cite{vollmayr+aspelmeier11,vannoije+ernst99}.

In the long wave length limit $k\to0$ it holds that $S_k,
c_k\to$~const.\ and $U_{\mathrm{L}}^2(kd) \to 1$,
$U^2_{\mathrm{T}}(kd)\to2/3$. In the long-time limit $t\to\infty$, we
thus find 
\numparts
\begin{eqnarray}
  \label{eq:14}
  m_{\mathrm T}(t\to\infty) &\simeq -M_{\mathrm{T}}[(D + \eta)t/d^2]^{-3/2},\\
  m_{\mathrm L}(t\to\infty) &\simeq -M_{\mathrm{L}}[(D+\Gamma)t/d^2]^{-3/2}
  e^{-c^2t/4(D + \Gamma)},\\
  m_{\rho}(t\to\infty) &\simeq M_{\rho}[(D+\Gamma)t/d^2]^{-1/2}
  e^{-c^2t/4(D + \Gamma)}.
\end{eqnarray}
\endnumparts
This is the central result of this contribution.  The evaluation of
$m_{\mathrm{T}}(t\to\infty)$ is simply a moment of a gaussian
integral. The types of integrals that are necessary for the evaluation
of $m_{\rho,\mathrm{L}}(t\to\infty)$ are discussed in
\ref{sec:some-integrals}. Away from the glass transition, $c^2/4(D +
\Gamma) \sim \Or(\omega_{\mathrm{E}})$, i.e., the contributions
$m_{\rho,\mathrm{L}}(t)$ decay on a short time scale
$\propto\omega_{\mathrm{E}}^{-1}$. The dominant asymptotic
contribution is thus $m(t\to\infty) = m_{\mathrm{T}}(t\to\infty)$.

The prefactors read explicitly 
\numparts
\begin{eqnarray}
  \label{eq:16}
  M_{\mathrm{T}} &= \frac{1}{486\sqrt\pi}\frac{(1+\varepsilon)^2}{4\varphi},\\
  M_{\mathrm{L}} &= \frac{1}{162\sqrt\pi}\frac{(1+\varepsilon)^2}{4\varphi}
  \frac{c^2}{D+\Gamma},\\
  M_{\rho} &= \frac{1}{1152\sqrt\pi}\frac{1+\varepsilon}{2\varphi}
  \frac{S_0(nc_0)^2}{\omega_{\mathrm{E}}^2d^2}\frac{c^4}{(D + \Gamma)^4}.
\end{eqnarray}
\endnumparts
Due to the nontrivial dependence of the viscosity, $\eta(\varphi,\varepsilon)$,
and the sound damping $\Gamma(\varphi,\varepsilon)$ on the coefficient of
restitution, $\varepsilon$, and on the density, $\varphi$, there is no
simple trend of $m_{\mathrm{T},\mathrm{L},\rho}$ with $\varepsilon$. A
reduction of the memory effects compared to fluid of elastic hard
spheres, however, can be expected.

From equation~(\ref{eq:37}), it follows that
\begin{equation}
  \label{eq:36}
  \psi(t\to\infty) \simeq M_{\mathrm{T}}[(D + \eta)t/d^2]^{-3/2}
  \propto t^{-3/2}.
\end{equation}
With this result I have answered both questions from the
introduction. We now know the value of the exponent $\alpha$ and which
of the possible couplings is relevant.

At high densities, close to the granular glass transition
\cite{kranz+sperl13}, the viscosity, $\eta$, is expected to be large
and the long-time tail will be strongly suppressed
\cite{franosch+goetze98}.

\section{Summary \& Perspectives}
\label{sec:concl--persp}

I discussed the coupling of the tagged particle velocity to the
hydrodynamic modes of the host fluid in the frame work of
mode-coupling theory. Considering a randomly driven inelastic hard
sphere fluid with local momentum conservation, I found that the VACF
decays algebraically, $\psi(t\to\infty) \propto t^{-\alpha}$, with an
exponent $\alpha=3/2$. This supports observations from simulations
\cite{fiege+aspelmeier09}. The relevant process for the algebraic
decay is, both for elastic and inelastic hard spheres, the coupling to
the transverse currents. The coupling to the density and longitudinal
currents have a finite life time.

The discussion of the VACF in a randomly driven granular fluid
\emph{without} momentum conservation will be left to future
work. This could possibly also help to settle the question about the
nature of the long-time tails in the sheared granular
fluid. 

\ack

I would like to express my gratitude to Annette Zippelius for
initiating this study and for long time support. I thank Andrea Fiege
and Matthias Sperl for many illuminating discussions and Matthias for
critically reading the manuscript.

\appendix

\section{Vertices}
\label{sec:appendix}

Here, I will detail the calculation of the vertices.  

\subsection{Longitudinal}
\label{sec:longitudinal}

Due to the symmetry of the velocity distribution function, we have
\begin{eqnarray}
  \label{eq:41}
  \avr{\vec v_s|\Lv_+\Q j_{-\vec k}^{\mathrm{L}}\rho_{\vec k}^{\mathrm{s}}}
  &= \avr{\vec v_s|\rho_{\vec k}^{\mathrm{s}}\mathcal T_+j_{-\vec k}^{\mathrm{L}}}
  - \frac13\avr{\vec v_s|\mathcal T_+\vec v_s}
  \avr{\vec v_s|j_{-\vec k}^{\mathrm{L}}\rho_{\vec k}^{\mathrm{s}}}\nonumber\\
  &= \uvec k\avr{j_{\vec k}^{\mathrm{sL}}|\mathcal T_+j_{\vec k}^{\mathrm{L}}}
  - \rmi\frac{1+\varepsilon}{3}\omega_{\mathrm{E}}\uvec k
  \avr{j_{\vec k}^{\mathrm{sL}}|j_{\vec k}^{\mathrm{L}}}
\end{eqnarray}
and
\begin{eqnarray}
  \label{eq:42}
  \avr{j_{-\vec k}^{\mathrm{L}}\rho_{\vec k}^{\mathrm{s}}\Q|\Lv_+\vec v_s}
  &= \avr{j_{-\vec k}^{\mathrm{L}}\rho_{\vec k}^{\mathrm{s}}|\mathcal T_+\vec v_s}
  - \frac13\avr{j_{-\vec k}^{\mathrm{L}}\rho_{\vec k}^{\mathrm{s}}|\vec v_s}
  \avr{\vec v_s|\mathcal T_+\vec v_s}\nonumber\\
  &= \uvec k\avr{j_{\vec k}^{\mathrm{L}}|\mathcal T_+j_{\vec k}^{\mathrm{sL}}}
  - \rmi\frac{1+\varepsilon}{3}\omega_{\mathrm{E}}\uvec k
  \avr{j_{\vec k}^{\mathrm{L}}|j_{\vec k}^{\mathrm{sL}}}.
\end{eqnarray}
This shows that the left and the right vertex are identical. With
$\avr{j_{\vec k}^{\mathrm{L}}|j_{\vec k}^{\mathrm{sL}}} = 1/N$ and
$\avr{j_{\vec k}^{\mathrm{L}}|\mathcal T_+j_{\vec k}^{\mathrm{sL}}} =
\nu_k/N$, where $\nu_k$ was determined in \cite{kranz+sperl13}, 
equation~(\ref{eq:25}) follows.

\subsection{Transverse}
\label{sec:transverse}

Starting like in the longitudinal case, we find
\begin{equation}
  \label{eq:8}
  \avr{\vec j_{-\vec k}^{\mathrm{T}}\rho_{\vec k}^{\mathrm{s}}\Q|\Lv_+\vec v_s}
  = \avr{\vec j_{\vec k}^{\mathrm{T}}|\mathcal T_+\vec j_{\vec k}^{\mathrm{sT}}}
  - \rmi\frac{1+\varepsilon}{3}\omega_{\mathrm{E}}
  \avr{\vec j_{\vec k}^{\mathrm{T}}|\vec j_{\vec k}^{\mathrm{sT}}}.
\end{equation}
The proof that the left and right vertices are identical is completely
analogous to the discussion above. 

We have $\avr{\vec j_{\vec k}^{\mathrm{T}}|\vec j_{\vec
    k}^{\mathrm{sT}}} = 2/N$ and $\avr{\vec j_{\vec
    k}^{\mathrm{T}}|\mathcal T_+\vec j_{\vec k}^{\mathrm{sT}}} =
\avr{\vec j_{\vec k}|\mathcal T_+\vec j_{\vec k}^{\mathrm{s}}} -
\avr{j_{\vec k}^{\mathrm{L}}|\mathcal T_+j_{\vec k}^{\mathrm{sL}}}$.
With $\vec v = (\vec v_1 - \vec v_s)/\sqrt2$ and $\vec r = \vec r_1 -
\vec r_s$ we write
\begin{equation}
  \label{eq:9}
  \avr{\vec j_{\vec k}|\mathcal T_+\vec j_{\vec k}^{\mathrm{s}}}
  = \rmi\frac{1+\varepsilon}{2}\sqrt2\avr{(\uvec r\cdot\vec
    v)^3\Theta(-\uvec r\cdot\vec v)\delta(r - d)\left(e^{-\rmi\vec
        q\cdot\vec r} - 1\right)}
\end{equation}
where $\sqrt2\avr{(\uvec r\cdot\vec v)^3\Theta(-\uvec r\cdot\vec v)} =
-2/\sqrt\pi$ and
\begin{eqnarray}
  \label{eq:10}
  \avr{\delta(r - d)\left(e^{-\rmi\vec q\cdot\vec r} - 1\right)}
  &= \frac{2\pi d^2\chi}{V}
  \int_0^{\pi}\rmd\vartheta\sin\vartheta\left(e^{-\rmi qd\cos\vartheta} -
    1\right)\nonumber\\
  &= -24\frac{\varphi\chi}{dN}[1 - j_0(qd)],
\end{eqnarray}
i.e.,
\begin{equation}
  \label{eq:11}
  \avr{\vec j_{\vec k}|\mathcal T_+\vec j_{\vec k}^{\mathrm{s}}}
  = 2\rmi\frac{1+\varepsilon}{2N}\omega_{\mathrm{E}}[1 - j_0(qd)].
\end{equation}
Using suitable relations between spherical Bessel functions, 
equation~(\ref{eq:12}) follows.

\section{Some Integrals}
\label{sec:some-integrals}

All the integrals needed for $m_{\rho,\mathrm{L}}(t\to\infty)$ can be
expressed as derivatives of 
\begin{equation}
  \label{eq:31}
  I(c, G; t) := \int_0^{\infty}\rmd k\cos(ckt)e^{-Gk^2t}
  = \frac12\sqrt{\frac{\pi}{Gt}}\exp(-c^2t/4G),
\end{equation}
where the second equality is given in \cite{gradshteyn00}. Then we
have
\begin{equation}
  \label{eq:32}
  \int_0^{\infty}\rmd kk^2\cos(ckt)e^{-Gk^2t} 
  = -\frac1t\frac{\partial I}{\partial G}
  = - \frac14I(c,G; t)\frac{c^2t - 2G}{G^2t},
\end{equation}
\begin{eqnarray}
  \label{eq:33}\fl
  \int_0^{\infty}\rmd kk^4\cos(ckt)e^{-Gk^2t} 
  = \frac{1}{t^2}\frac{\partial^2I}{\partial G^2}\nonumber\\
  = \frac{1}{16}I(c, G; t)\frac{12G^2 - 12c^2Gt + c^4t^2}{G^4t^2},
\end{eqnarray}
and,
\begin{equation}
  \label{eq:34}
  \int_0^{\infty}\rmd kk^3\sin(ckt)e^{-Gk^2t}
  = \frac{1}{t^2}\frac{\partial^2I}{\partial G\partial c}
  = \frac18I(c, G; t)c\frac{c^2t - 6G}{G^3t}.
\end{equation}
In particular,
\begin{equation}
  \label{eq:35}
  2cG\frac{1}{t^2}\frac{\partial^2I}{\partial G\partial c}
  + c^2\frac1t\frac{\partial I}{\partial G} 
  = \frac{\sqrt\pi}{2}\frac{c^2}{G}(Gt)^{-3/2}\exp(-c^2t/4G).
\end{equation}

\section*{References}
\label{sec:references}

\bibliography{literatur}

\end{document}